\documentclass[acmsmall]{acmart} %

\AtBeginDocument{%
  \providecommand\BibTeX{{%
    \normalfont B\kern-0.5em{\scshape i\kern-0.25em b}\kern-0.8em\TeX}}}

\begin{document}

\title{``Alexa, Can I Program You?'': Student Perceptions of Conversational %
Artificial Intelligence Before and After Programming Alexa}

\author{Jessica Van Brummelen}
\email{jess@csail.mit.edu}
\orcid{0000-0002-4831-6296}
\author{Viktoriya Tabunshchyk}
\email{vikt@mit.edu}
\author{Tommy Heng}
\email{theng@mit.edu}
\affiliation{%
  \institution{Massachusetts Institute of Technology}
  \streetaddress{77 Massachusetts Ave}
  \city{Cambridge}
  \state{MA}
  \country{USA}
  \postcode{02139}
}

\renewcommand{\shortauthors}{Van Brummelen, et al.}

\begin{abstract}
  Growing up in an artificial intelligence-filled world, with Siri and Amazon Alexa often within arm's---or speech's---reach, could  have significant impact on children. Conversational agents could influence %
  how students anthropomorphize computer systems or develop a theory of mind. Previous research has explored how conversational agents are used and perceived by children within and outside of learning contexts. %
  This study investigates how middle and high school students' perceptions of Alexa change through programming their own conversational agents in week-long AI education workshops. Specifically, we investigate the workshops' influence on student perceptions of Alexa's intelligence, friendliness, aliveness, safeness, trustworthiness, human-likeness, and feelings of closeness. We found that students felt Alexa was more intelligent and felt closer to Alexa after the workshops. We also found strong correlations between students' perceptions of Alexa's friendliness and trustworthiness, and safeness and trustworthiness. Finally, we explored how students tended to more frequently use computer science-related diction and ideas after the workshops. 
  Based on our findings, we recommend designers carefully consider personification, transparency, playfulness and utility when designing CAs for learning contexts.
\end{abstract}



\begin{teaserfigure}
  \includegraphics[width=\textwidth]{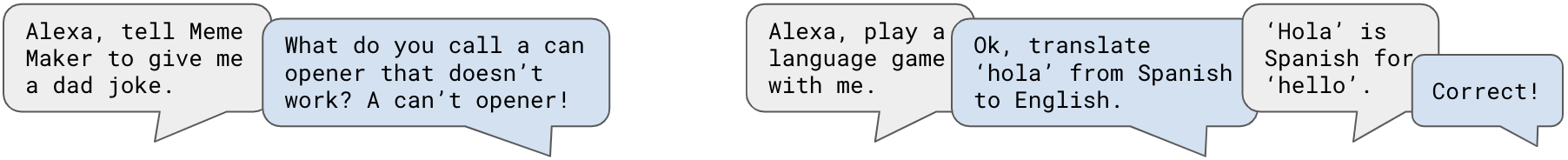}
  \caption{Example conversations from students' Alexa skill designs, including a ``Meme Maker'' and ``Language Game''.}
  \Description{Speech bubbles of conversations from students' Alexa skill designs. The ``Meme Maker'' conversation reads, ``Alexa tell Meme Maker to give me a dad joke.'' ``What do you call a can opener that doesn't work? A can't opener!'' The ``Language Game'' conversation reads, ``Alexa, play a language game with me.'' ``Ok, translate `hola' from Spanish to English.'' ```Hola' is Spanish for `hello'.'' ``Correct!''}
  \label{fig:teaser}
\end{teaserfigure}

\maketitle

\section{Introduction}
With children asking Google to buy them more toys \citep{alexa-kids-twitter}, cheating on homework with Alexa \citep{alexa-kids-homework}, %
and playing voice-based pranks on parents \citep{alexa-kids-twitter}, 
conversational agents (CAs) have potential to not only influence children's play---but also how they grow and develop \citep{whom-linguistic-spitale}. For instance, researchers theorize that interacting with an agent can change people's understanding of agency concepts and their Theory of Mind (ToM) %
\citep{agency-teachable-agent-jaeger,agency-human-robot-levin,robot-ToAM-spektor}. Other research has shown engaging with CAs can change people's behavior \citep{conv-ai-persuiasion-adler,conv-ai-repair-misunderstandings-corti,socially-desirable-conv-ai-scheutzler,ai-influence-cognition-qian} and have positive effects on information retention \citep{conv-agent-memory-beun}.

Considering the impact agents have on human understanding and behavior, how prevalent these systems are becoming \citep{whom-linguistic-spitale,conv-ai-market}, and how opaque their operations can be to humans \citep{advocacy-ai-literacy-register,hey-google-eat-druga,hey-google-unicorns-lovato}, 
a growing body of research suggests it is important for people of all ages to understand AI \citep{ai-literacy-long,big-ai-ideas-touretzky,advocacy-ai-literacy-register}. Furthermore, researchers are investigating how to best teach AI literacy concepts to students, including those as young as preschoolers \citep{popbots-robot-perceptions-randi}. For instance, one study leverages 3-5th grade students familiarity with CAs to teach AI literacy concepts \cite{zhorai}. Other works utilize AI ethics discussions \citep{ai-ethics-daniella-blakeley}, interactive, collaborative learning environments \citep{smiley}, and gesture recognition tools \citep{gesture-zimmermann} to engage students in learning AI. In this work, we use a constructionist approach, in which students program their own CAs, to teach AI concepts to 6-12th grade students \citep{appinv-convai-eaai,constructionism-papert}.

Another aspect of AI education research includes students' perception of AI systems themselves, including personification of such systems, emotions the systems evoke, and students' conceptions of how the systems work. For example, one study examines preschool- and kindergarten-aged students' perceptions of ``thinking machines'' during an AI learning activity, emphasizing the importance of early childhood AI literacy and ToM development \citep{popbots-robot-perceptions-randi}. Other studies investigate children and family's perceptions of CAs \citep{hey-google-unicorns-lovato}, how interaction modalities influence children's perceptions of CAs \citep{hey-google-eat-druga}, children's perceptions of maze-solving agents' intelligence \citep{how-smart-toys-druga}, and whether children categorize CAs as animate objects or artifacts \citep{what-you-talking-children-xu}. Yet other studies emphasize the importance of adults' perceptions and conceptions of AI, especially in decision-making and policy %
\citep{asimo-policy-law-ToM-jaeger,drivers-car-conv-ai-large,computer-identity-impressions-shank}. %
To our knowledge, few studies investigate middle and high school students' perceptions of AI \citep{learningml-rodriguez,supporting-science-robot-michaelis}, despite teenage years being critical in ethical perspective development \citep{youth-development-ethics-damon}, a key component of AI literacy \citep{big-ai-ideas-touretzky}. %
Furthermore, to our knowledge, no studies investigate how middle and high school students' perceptions of CAs change through programming CAs.

We posit that understanding students' perceptions and feelings towards such agents can help researchers better facilitate student learning. %
For instance, feelings of closeness with teachers have been shown to affect students' academic performance \citep{teacher-child-closeness-adjustment-birch,gender-child-teacher-closeness-wolter,socioemotional-closeness-attachment-al-yagon}, which may also be the case when agents take on the teacher role. Another study indicates that the avatar used for pedagogical feedback-giving agents affect students' emotional attachment and satisfaction with the learning process \citep{avatar-digital-learning-schobel}, alluding to the potential for students' perception of agents to affect learning. Furthermore, research suggests understanding students' preconceptions and mental models can improve teaching \citep{conceptions-science-teaching-sherin,bibliography-conceptions-duit}. By understanding students' feelings and conceptions %
about agents, we expect we can create better digital learning environments.

This study investigates 6-12th grade students' perceptions and conceptions of Amazon Alexa in a learning environment described in \citep{appinv-convai-eaai}. In contrast to \citep{appinv-convai-eaai}, which investigates students' AI literacy, this study investigates %
how a programming and learning intervention, in which students develop their own CAs, affects student perspectives of AI. %
Our main research question is as follows:

\textbf{RQ: How does building Alexa skills and learning about conversational AI in a remote workshop affect students' perceptions and conceptions of AI, conversational AI, and Alexa?}

By better understanding students' perspectives on agents and how these perspectives can be changed, we contribute to ongoing research to develop more human-centered, socially useful agents---especially for K-12 education. To this end, we present four design considerations for K-12 education agents and development tools based on our findings. Specifically, we look at students' conceptions of how AI and conversational AI work, %
and perceptions of Alexa in terms of friendliness, human-likeness, aliveness, safeness, trustworthiness, intelligence (generally and relative to themselves), and how close they feel to Alexa.

\section{Related Work}

\subsection{Conceptions of artificial intelligence and theory of mind}

The working definition of AI in research has changed over the years---from having a sharp focus on logical, symbolic representations of concepts and actions to a marked concentration on modelling extensive interconnected computation machines called ``neural networks'' \citep{ai-history-wooldridge}. In the media, AI has been depicted in many different ways---as killer robots, android caretakers, and superintelligent, disembodied voices \citep{androids-ai-media-guadamuz}. Despite the somewhat frivolous portrayals, people's understanding of AI and how it works has serious implications---from policy-making to day-to-day assessments of whether a self-driving vehicle is safe to trust one's life with \citep{asimo-policy-law-ToM-jaeger,autonomous-vehicles-chi-workshop}.

ToM research in AI investigates how to develop AI systems with human-like cognition, as well as how people understand AI as agents with mental states \citep{ai-ToM-erb}. In this research, we focus on the latter, or ``Theory of Artificial Mind'' (ToAM) \citep{robot-ToAM-spektor}. Understanding people's conceptions of AI, including anthropomorphization of AI technology, conceptions of specific technologies, like CAs, and emotional reactions to AI systems, is important for teaching AI literacy. %
Through better understanding students' perceptions and ToAM, we can likely better teach students about AI \citep{agency-teachable-agent-jaeger} and therefore better reach our research community's goal of equipping people to live in an AI-filled world \citep{asimo-policy-law-ToM-jaeger,big-ai-ideas-touretzky,ai-literacy-long}.

Children have been observed to anthropomorphize AI systems \citep{how-smart-toys-druga,hey-google-eat-druga,what-you-talking-children-xu}; however, their understanding of the actual ``aliveness'' of such systems is inconsistent across populations and seems to vary with age \citep{popbots-robot-perceptions-randi,computers-brains-children-scaife,robovie-closet-children-kahn,what-you-talking-children-xu}. Other anthropomorphic aspects of AI systems have also been investigated for different purposes. For instance, a number of studies examine how children (3-10 years old) perceive agents' intelligence---generally and relative to their intelligence---with the purpose of inspiring critical thinking \citep{how-smart-toys-druga,hey-google-eat-druga}. Another study investigates 5-6 year old children's perception of CAs' friendliness, aliveness, trustworthiness, safeness, and funniness, in addition to intelligence to develop CA design recommendations \citep{hey-google-unicorns-lovato}. Researchers investigated similar anthropomorphic aspects, including how sociable, mutual-liking, attractive, human, close, and intelligent children (10-12 years old) perceive agents to be, in order to improve learning interventions \citep{supporting-science-robot-michaelis}. We investigate related anthropomorphic aspects in middle to high school students' ToAM.

Research also shows that interaction with AI artifacts can influence people's ToM and perceptions of AI. For instance, observing and constructing robot behavior influenced students' ToAM, enabling them to better explain the AI systems' behavior \citep{robot-ToAM-spektor}. Another study showed that interacting with a pedagogical agent influenced students' understanding of the key ToM concept of agency, allowing them to better predict behavior. The same study linked students' prior understanding of agency to better learning \citep{agency-teachable-agent-jaeger}. In this paper, we investigate how AI literacy workshops involving programming a CA influences students' ToAM, including perceptions of anthropomorphic qualities and understanding of AI behavior. %

\subsection{Conversational agents and education}

Many studies investigate how CAs can best embody the teaching role \citep{multiple-agents-academic-success-dincer,emotion-pedagogic-conv-ai-morales,exploratory-interact-pedagogic-perez,bettys-brain-og-leelawong}. Some such studies show that interacting with agents can positively affect learning and students' ToM \citep{multiple-agents-academic-success-dincer,agency-teachable-agent-jaeger}. In this study, however, we take a constructionist approach, and instead of placing agents in the teaching role, we empower students to learn about AI through developing their own CAs \citep{constructionism-papert,appinv-convai-eaai}.

Constructionism has been shown to be effective in teaching K-12 students AI concepts. For example researchers have taught students AI ethics through constructing paper prototypes %
\citep{constructionism-ethics-safinah-blakeley}, machine learning (ML) concepts through developing gesture-based models \citep{gesture-zimmermann}, and AI programming concepts through creating projects with AI cloud services \citep{snap-ai-programming-kahn}. Our study teaches students \citeauthor{ai-literacy-long}'s AI literacy competencies through developing CAs \citep{appinv-convai-eaai,ai-literacy-long}.

Certain studies specifically investigate whether constructionist activities change student conceptions and perceptions of AI agents. For example, a series of studies showed constructing a robot's behavior enabled kindergarten students to conceptualize an agent's rule-based behavior  \citep{constructing-robot-behavior-mioduser}, shifted students' perspectives from technological to psychological \citep{want-programmed-robots-levy}, and shifted students' language from anthropomorphic to technological \citep{anthropomorphic-artifacts-behavior-kuperman}. Through an activity with the same constructionist programming environment, it was shown 5- and 7-year-old students' conceptions of %
ToAM developed, and the students were able to better understand robots' behavior \citep{robot-ToAM-spektor}. %
A study with programming and ML training activities showed 4-6-year-old students' understanding of ToM and perceptions of robots changed throughout the experiment \citep{popbots-robot-perceptions-randi}. In this work, we investigate whether students' ToAM and perceptions of AI in middle and high school change through a constructionist CA programming activity and workshop.

\section{Methods}

\subsection{Participants}
We conducted our workshops with 47 students separated into two groups of 12 and 35. For each group, the students' teachers observed the workshops and provided feedback to the three teaching researchers \citep{appinv-convai-eaai}. %
The teachers were recruited through an Amazon Future Engineers call to Title I schools. Each teacher chosen for the workshops was asked to recruit 5 or 6 of their students. We targeted Title I schools because they have high concentrations of children from low-income families \citep{title-I-schools}, and we wanted to provide opportunities for enrichment that they may not normally receive. We developed middle and high school level AI curriculum and thus targeted middle and high school students. The students' mean age was 14.78 (range 11-18, SD=1.91), with 19 self-identifying as male, 27 self-identifying as female, and 1 student that did not complete the questionnaire. %

\subsection{Procedure}

\subsubsection{Programming agents}

To accomplish our goal of studying student perceptions of conversational AI before and after programming Alexa, we developed an interface within MIT App Inventor for creating Alexa skills \citep{vanbrummelen-sm}. This lowered the barrier of entry to programming CAs, as the students could use visual block-based coding to develop skills. 
As described in \citep{appinv-convai-eaai}, once a student creates a skill on the interface, the backend translates their blocks into JSON and Javascript code to be sent to Alexa's API to build and enable the skill on the student's Amazon Developer account. This allows the students to interact and have a conversation with the Alexa skill either on an Alexa-enabled device (e.g., iOS Alexa App or Amazon Echo) or an online simulated Alexa device (e.g., MIT App Inventor Alexa Testing simulator or Amazon Developer Console).

\subsubsection{Workshop outline}

This section provides a brief overview of the learning intervention, which is described in-depth in \citep{appinv-convai-eaai}. The intervention occurred over two sessions, which both involved five consecutive days %
of 2.5 hour long Zoom sessions. The first day began with an introduction to the MIT App Inventor interface \citep{appinv} to accustom students to block-based coding. Then the students were given a chance to interact freely with Alexa, writing down the questions they asked during the interaction. In the first week, students were each provided with a complimentary Echo Dot. This was not feasible for the second week of workshops due to an increased number of students, so students either used the Alexa app on their mobile devices, an online Alexa simulator (within MIT App Inventor or otherwise), or Alexa devices they previously owned. Overall, 19 students used an Alexa device, 17 used the Alexa app, 10 used an online simulator, and one did not specify.

The second day involved introducing students to key AI and conversational AI concepts, discussing AI ethics, and completing a tutorial walk-through to create an Alexa skill that would respond to basic greetings. %
On the third day, students completed a tutorial to develop a calculator skill, in which Alexa could be asked, ``What's \texttt{number A} multiplied by \texttt{number B}'', or something similar. %
Next, we taught students about ML in more depth, including discussing the difference between a rule-based CA developed on the first day and the ML-based CAs developed on the second and third days. Finally, students engaged in an AI text generation activity. %

On the fourth day, students developed a skill that enabled Alexa to read out text entered into MIT-App-Inventor-developed mobile apps. Students then brainstormed ideas for skills for their personal projects. Students spent the final day developing their projects and presenting them to the rest of the class.

\subsection{Questionnaires}

Various questionnaires inspired by the perception of AI questions in \citep{how-smart-toys-druga} and \cite{hey-google-unicorns-lovato} were given to students during the learning intervention. %
On the first day, students recorded their interactions with Alexa, impressions of the CA, and demographics information.
At the start of the second day, students completed a questionnaire assessing their initial feeling towards and understanding of Alexa, AI and conversational AI. %
The questions were divided into two sets, which we refer to as the \textit{Persona} and \textit{Conception} questions. 

The \textit{Persona} questions assessed students' sentiments about Alexa on a 7-point Likert scale. The questions stated, ``Alexa is...'' followed by ``intelligent'', ``friendly'', ``alive'', ``safe'', ``trustworthy'', ``human-like'', and ``smarter than me''. The final \textit{Persona} question asked how close students felt to Alexa using the \textit{Inclusion of the Other in the Self} scale \citep{closeness-question-gachter}. %
The \textit{Conception} questions assessed students' understanding of AI and conversational AI %
through %
asking, ``Describe in your own words what AI is'' and ``Describe in your own words what conversational AI is (e.g., chatbots, like Alexa or Google home, use conversational AI)''.
At the end of the final day, students completed the \textit{Persona} and \textit{Conception} questions again. %
Additional questionnaires were given at the end of the second, third, and fourth days to assess specific AI literacy competencies, as discussed and analyzed in \citep{appinv-convai-eaai}.

\subsection{Data Analysis}

This study builds on the study presented in \citep{appinv-convai-eaai}. Thus, certain data analyzed in this study (e.g., demographics) is necessarily the same; however, this study focuses on data not analyzed in \citep{appinv-convai-eaai}, including the questionnaire responses to the \textit{Persona} questions and students' reported interactions with Alexa. %
The responses to the \textit{Conception} questions were analyzed in both studies, however using different methods and through different lenses. This study investigates students' conceptions of AI through a word frequency analysis as well as analyses of changes in number of tags (as described below). The study in \citep{appinv-convai-eaai} assessed students' AI literacy before and after the learning intervention.

To investigate the responses to qualitative questions, a reflexive, open-coding approach to thematic analysis \citep{thematic-analysis-open-coding} was performed by three researchers. The three researchers independently completed familiarization and code-generation stages. After several discussions, the three researchers came to a consensus on codes for the questionnaire responses. Codes and respective representative quotations can be found in \citep{gist-appendix}. Researchers generally constructed codes inductively or with respect to ideas from literature, including the Big AI Ideas \citep{big-ai-ideas-touretzky}. It is important to note that responses often involved multiple ideas and were thus tagged with more than one code.

For the quantitative questions (e.g., Likert scale \textit{Persona} questions) asked on both pre- and post-questionnaires, the Wilcoxon Signed-Rank Test was employed to measure changes. Additionally, we used the Kendall Tau method to create pairwise correlation matrices. We analyzed the correlation coefficients using \citeauthor{correlation-coeff-cohen} (\citeyear{correlation-coeff-cohen})'s definition for correlation effect strength for behavioral and education psychology \citep{correlation-coeff-cohen}. To test the validity of the strength of the coefficients, we compared Kendall Tau p-values to an alpha of 0.05.

For the word frequency analysis, we used the NLTK library \citep{nltk} to remove stop-words, tokenize and lemmatize qualitative responses. Additionally, to better visualize non-obvious concepts, we filtered out words directly from the questions, including `AI', `artificial', `intelligence', and `conversational'. Word clouds were generated using \citep{wordcloud}.

\section{Results}\label{sec:results}

\subsection{Student interactions with Alexa}
To understand the types of interactions students had with Alexa prior to the intervention, we coded the phrases they reported saying to Alexa during the interaction activity. We found most of the phrases fell into one of five categories listed in Tab. \ref{tab:ques-asked}. %
The \textit{Information Updates} category involved real-time events; the \textit{Action Commands} category involved built-in Alexa applications; the \textit{Personal Questions} category involved questions about Alexa; the \textit{Jokes} category involved asking Alexa to say a joke; and the \textit{Other} category involved questions and phrases that were often humorous (e.g., ``Are dragons real?'') or impossible to fully answer (e.g., ``What are all the numbers of pi?''), or generally fell outside of the other categories (e.g., ``Hello''). %
Note that prior to the activity, we asked Alexa to tell us a joke, which may have contributed to a large number of students also asking Alexa for jokes.

\begin{table}
\small
\centering
\caption{Types of questions asked by students to Alexa prior to the conversational AI programming intervention.}
\label{tab:ques-asked}
\begin{tabular}{p{0.19\linewidth} | p{0.65\linewidth} | p{0.09\linewidth}}
\textbf{Type}       & \textbf{Example utterances}                                                                                                                                     & \textbf{Instances}  \\ 
\hline
Information updates & What time is it?, How is the weather for Wednesday?, How is the traffic? %
& 31 (26\%)           \\
Action commands     & Set a 15-minute timer, Play my Custom Spotify Playlist, Remind me that I have a meeting at 1:00 pm, What's 0 times 0? & 30 (25\%)           \\
Other               & Hello, Learn my voice, Are dragons real?, What are all the numbers of pi?                                                                    & 24 (20\%)           \\
Jokes               & Tell me a joke, Can you tell me a joke?                                                                                                                         & 17 (14\%)           \\
Personal questions  & What's your favorite color?, When were you made?, What's your favorite video game?, How was your day?                                                           & 16 (14\%)          
\end{tabular}
\end{table}

\subsection{Perceptions of Alexa pre- and post-workshop}
\begin{figure}[htb!]
    \centering
	\includegraphics[width=0.6\linewidth]{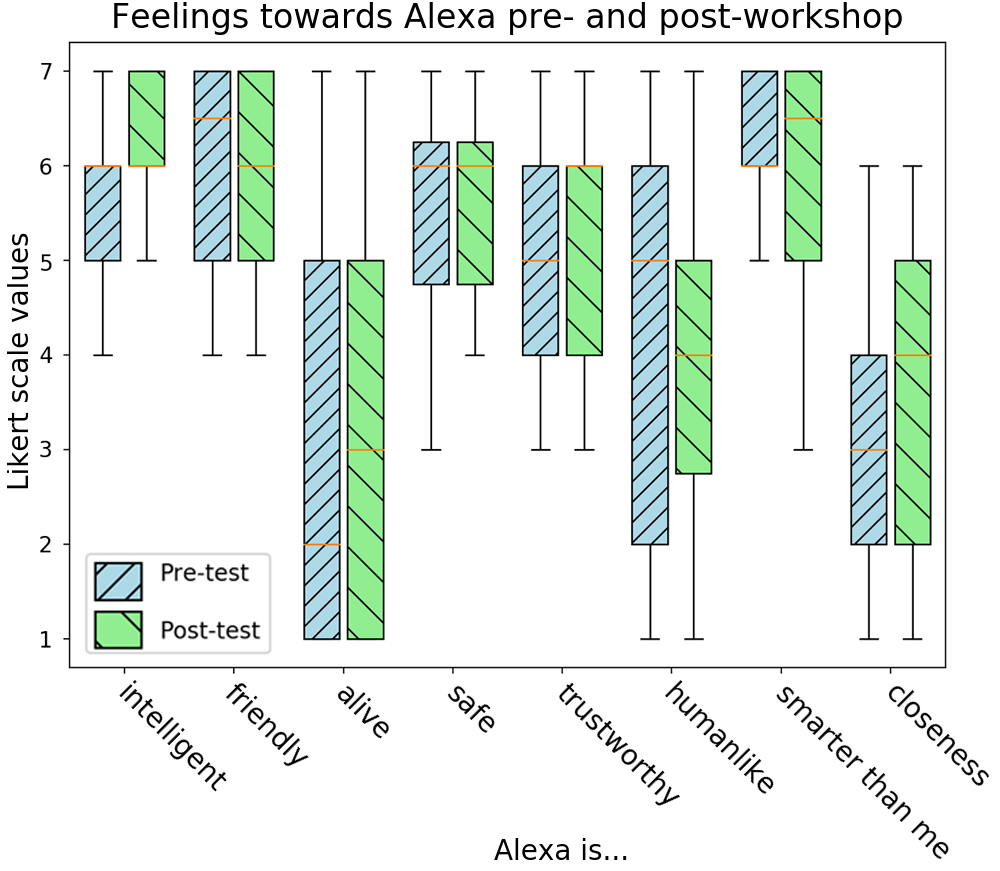}
	\caption{Students' perceptions of Alexa prior to the workshops (in blue) and after (in green).}
	\label{fig:likert-persona}
\end{figure}

By comparing pre- and post-survey answers to the \textit{Persona} questions (see Fig. \ref{fig:likert-persona}), we found significant differences in how students felt about Alexa's intelligence and how close they felt they were to Alexa. After the intervention, students felt Alexa was more intelligent ($\bar x = 6.0$, $Mo = 6$, $|Z| = 2.78$, $p = 0.003$) and felt closer to Alexa ($\bar x = 3.5$, $Mo = 4$, $|Z| = 2.75$, $p = 0.003$). We did not find any evidence of significant differences in how students felt about Alexa being friendly, alive, safe, trustworthy, human-like or smarter than themselves before and after the intervention.

Prior to the intervention, students generally reported Alexa as being highly intelligent ($\bar x = 5.6$, $Mo = 6$), highly friendly ($\bar x = 6.0$, $Mo = 7$), not very alive ($\bar x = 2.9$, $Mo = 1$), highly safe ($\bar x = 5.4$, $Mo = 6$), moderately to highly trustworthy ($\bar x = 5.3$, $Mo = 4, 5$), and moderately human-like ($\bar x = 4.2$, $Mo = 5$). They also reported feeling Alexa was much smarter than themselves ($\bar x = 6.1$, $Mo = 7$), and feeling not particularly close to Alexa ($\bar x = 3.0$, $Mo = 2$). The results were similar after the intervention (other than the changes in %
intelligence and %
closeness described above).%

\subsection{Correlations between perceptions of Alexa}\label{sec:correlations}
We found strong ($r \geq 0.5$ \citep{correlation-coeff-cohen}) correlations between student reports of Alexa's safeness and trustworthiness on both the pre- and post-test, as well as between Alexa's friendliness and trustworthiness on the post-test. There was also a strong correlation between trustworthiness reported on the pre-test and safeness reported on the post-test. Student reports of Alexa's friendliness and trustworthiness on the pre-test and between the pre- and post-tests were moderately ($r \geq 0.3$ \citep{correlation-coeff-cohen}) correlated.

\begin{figure}[htb!]
    \centering
		\includegraphics[width=0.8\linewidth]{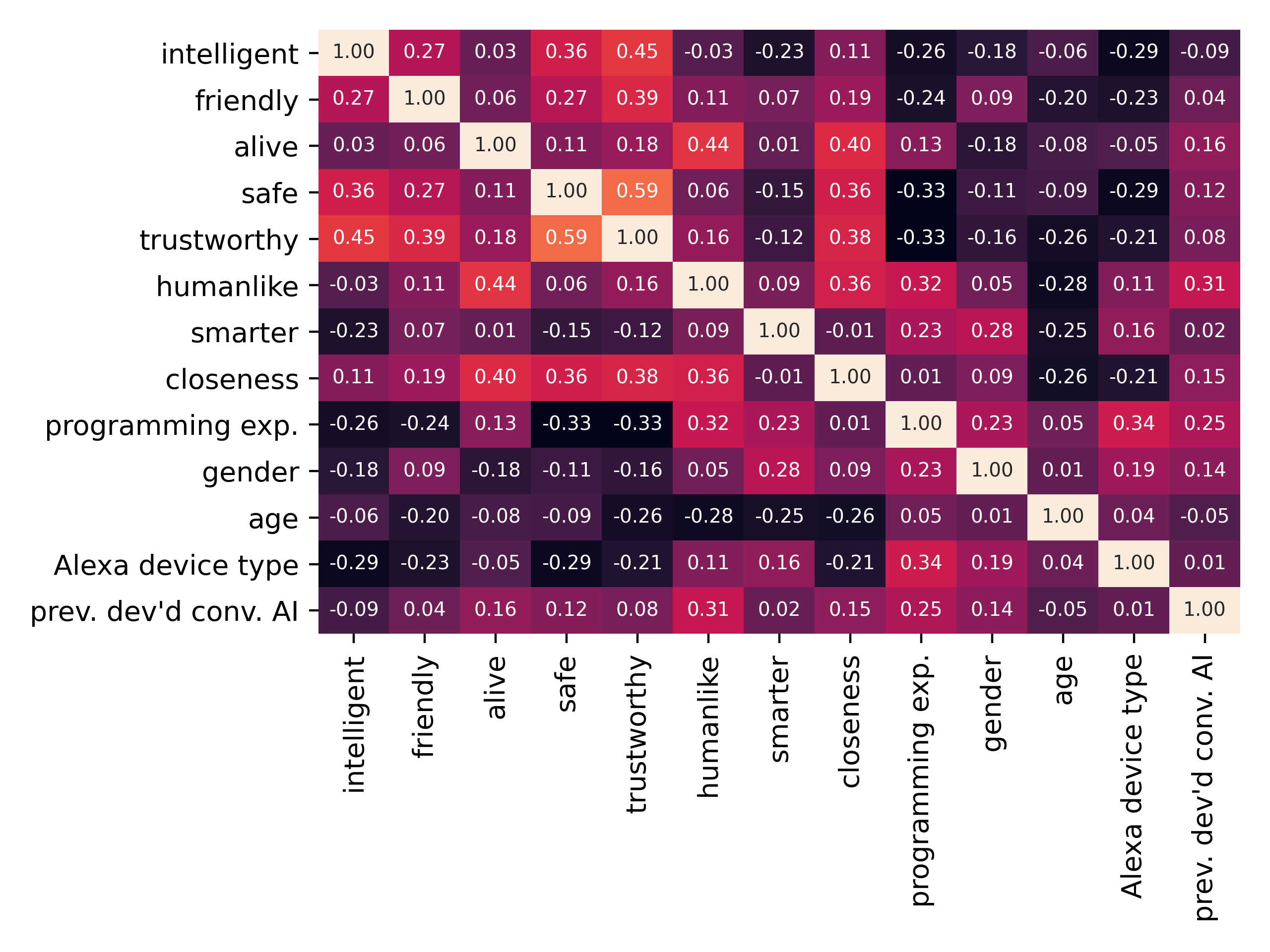}
		\includegraphics[width=0.8\linewidth]{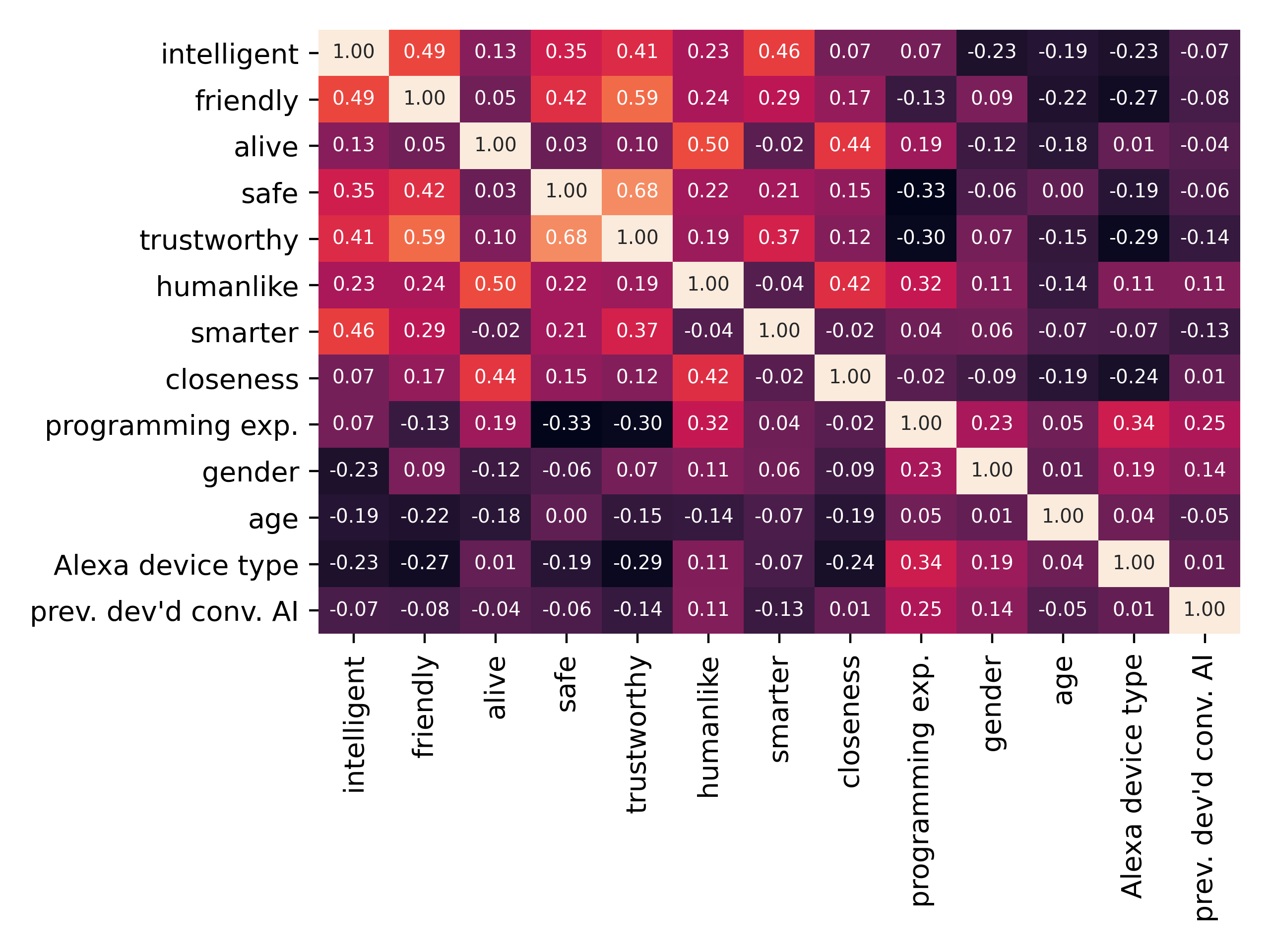}
	\caption{Correlation matrices before (top) and after (bottom) the intervention. Lighter colors correspond to higher coefficients.}
	\label{fig:correlations}
\end{figure}

Other moderate correlations included student reports of Alexa's intelligence and trustworthiness, friendliness and safeness, trustworthiness and feelings of closeness, human-likeness and aliveness, human-likeness and feelings of closeness, as well as aliveness and feelings of closeness. In the post-test, student reports of Alexa's intelligence and feeling Alexa was smarter than them, as well as Alexa's trustworthiness and feeling Alexa was smarter than them were moderately correlated. Additionally, there was a moderate correlation between students with more experience programming prior to the intervention and reports of Alexa's human-likeness on both the pre- and post-test. Our full correlation analysis is shown in Fig. \ref{fig:correlations}.

\subsection{Student diction when describing AI}
\begin{figure}[htb!]
    \centering
	\includegraphics[width=0.52\linewidth]{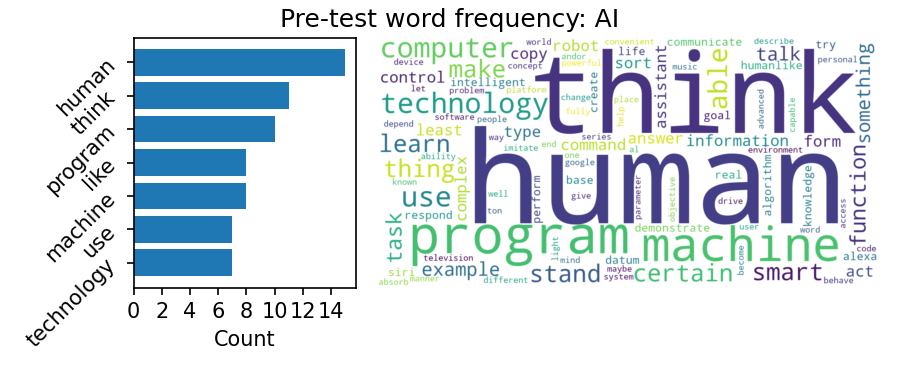}\includegraphics[width=0.52\linewidth]{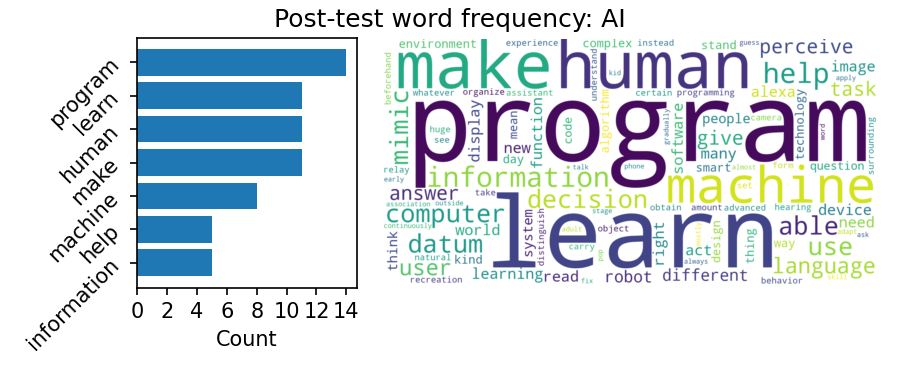}
	\caption{Word frequency analyses from %
	``Describe in your own words what AI is'' prior to (left) and after the intervention (right).}
	\label{fig:wordcloud-ai}
\end{figure}

\begin{figure}[htb!]
    \centering
	\includegraphics[width=0.52\linewidth]{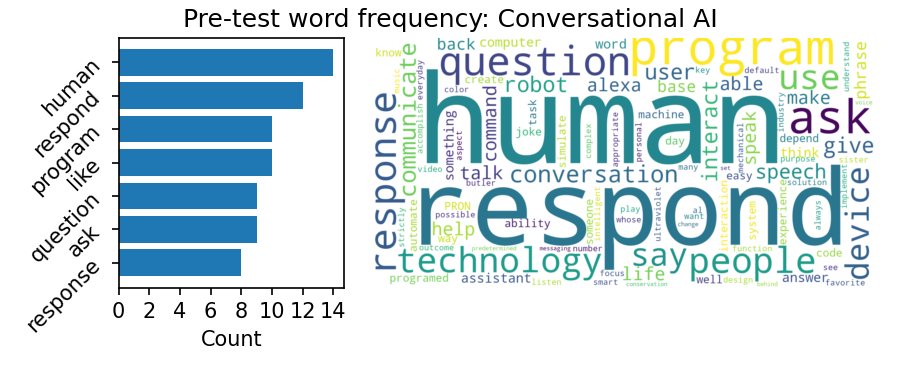}\includegraphics[width=0.52\linewidth]{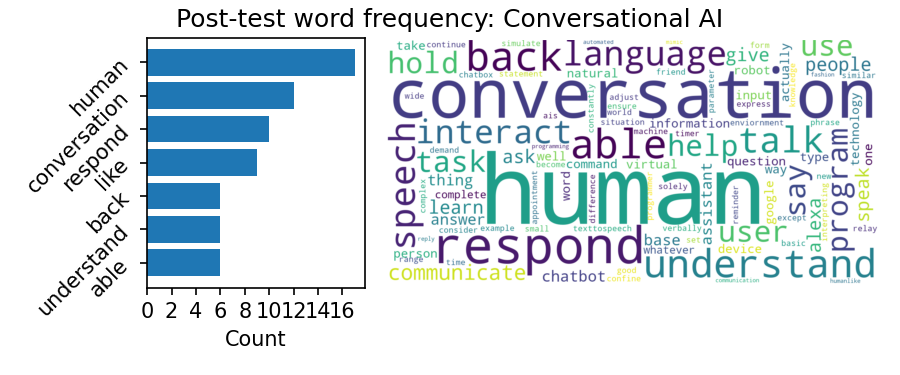}
	\caption{Word frequency analyses from %
	``Describe in your own words what conversational AI is'' prior to (left) and after the intervention (right). Notice the relative increase in the words, \textit{conversation}, \textit{back}, and \textit{able}, likely having to do with agents' abilities to have back and forth conversation.}
	\label{fig:wordcloud-conv-ai}
\end{figure}

To visualize students' understanding of AI and conversational AI, we analyzed word frequency and created word clouds based on answers to two questions. Fig. \ref{fig:wordcloud-ai} shows the word frequency analyses of students' answers to, ``Describe in your own words what AI is'', prior to and after the intervention. Fig. \ref{fig:wordcloud-conv-ai} shows the analyses of answers to, ``Describe in your own words what conversational AI is (e.g., chatbots, like Alexa or Google Home, use conversational AI)''.

\subsection{Conceptions of AI and conversational AI}
\begin{figure}[htb!]
    \centering
	\includegraphics[width=1\linewidth]{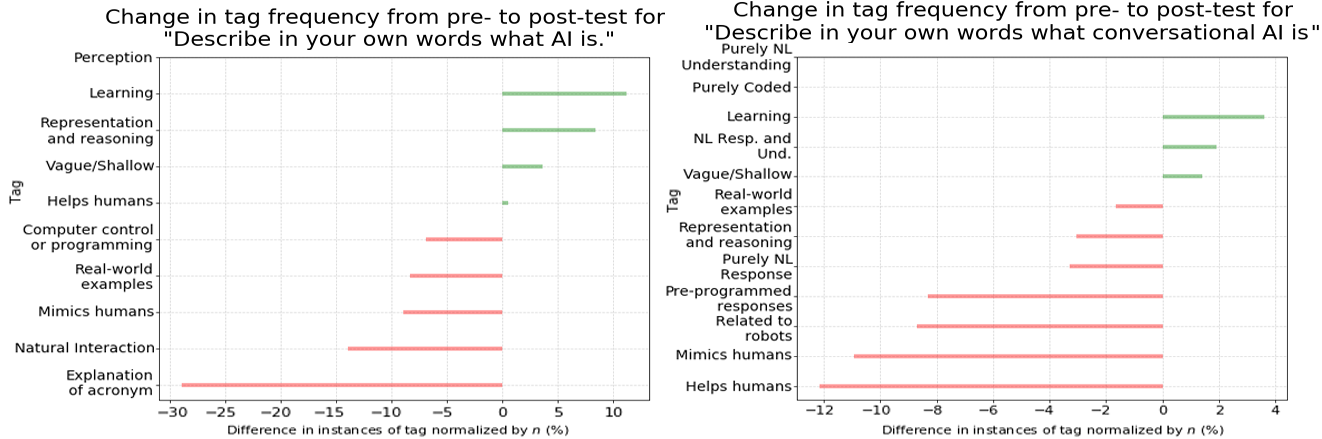}
	\caption{The change in instances of tags from pre- to post-test for \textit{Conception} questions. The change was calculated according to Eq. \protect\ref{eq:change}.}
	\label{fig:tags-conception}
\end{figure}

To better understand students' qualitative answers when conceptualizing AI and conversational AI, we performed a graphical exploration of tag frequency from our thematic analysis. %
The graph in Fig. \ref{fig:tags-conception} shows the change in tag frequency from pre- to post-test. Since the number of participants who completed the pre-test differed from the post-test, the number of tags ($t_{pre}$ and $t_{post}$) were normalized over the number of responses ($n_{pre} = 45$ and $n_{post} = 38$), and reported as normalized percent change, $C_\%$ (Eq. \ref{eq:change}). This is presented as an exploratory, graphical analysis for high-level insights rather than statistical analysis.

\begin{equation}
  C_\% = \frac{100*t_{post}}{n_{post}} - \frac{100*t_{pre}}{n_{pre}}\label{eq:change}
\end{equation}

\section{Discussion}
\subsection{Perceptions of Alexa's persona}
Prior to the study, we hypothesized students would feel Alexa was \emph{less} intelligent after learning how to program it, as they would better understand how it works; however, students felt Alexa was \emph{more} intelligent after the intervention ($|Z| = 2.78$, $p = 0.003$). This could have been for multiple reasons. Perhaps by successfully learning fundamental AI literacy concepts \citep{appinv-convai-eaai}, students realized Alexa was more complex than they initially thought and thus perceived it to be more ``intelligent'' (as in the Dunning-Kruger effect \citep{dunning-kruger-effect-olson}). This is supported by the relative increase in AI literacy concepts (which are comparatively complex) in the post-test responses to the \textit{Conception} questions (Fig. \ref{fig:tags-conception}), and the relative decrease in pre-programming concepts (which are comparatively simplistic). %
Students also generally felt Alexa was smarter than themselves (both before %
and after %
the intervention). This is consistent with previous studies of students aged 3-10 \citep{hey-google-eat-druga, hey-google-unicorns-lovato}. %

The Dunning-Kruger concept may also explain why there were relatively fewer tags identified---likely indicating fewer ideas presented by students---in the post-test than in the pre-test for many of the qualitative answers to the  %
\textit{Conception} questions. For example, as shown in Fig. \ref{fig:tags-conception}, there were relatively fewer tags %
for the majority of the tag categories in the post-test responses about conversational AI.  %
Perhaps students became ``less ignorant of their ignorance'' \citep{dunning-kruger-effect-olson} about Alexa through the intervention, and therefore felt less qualified to answer the qualitative questions and thus presented fewer ideas in the post-test. Nevertheless, one limitation of this study was that students responded to the post-test at the end of the workshops, so they may have had less energy than when they responded to the pre-test, alternatively explaining the relatively fewer ideas presented.

We also hypothesized that students would personify Alexa less after understanding the logic behind how it works, and therefore rate its ``aliveness'', ``human-likeness'', ``friendliness'', and their feelings of closeness to it as less than prior to the intervention. However, there was no significant evidence for any change, except that they felt closer to Alexa ($|Z| = 2.75$, $p = 0.003$) after the intervention. Students' increased feelings of closeness could be due to %
``boundary dissolution'', which is a type of closeness where two agents (usually human) no longer function completely autonomously, but rather function dependently \citep{closeness-examples-kreilkamp}. In this case, an apparent ``boundary dissolution'' due to Alexa initially seeming to function independently, but seeming to function dependently on students' programming efforts after the intervention, could have caused students' increased feelings of closeness.

Alternatively, perhaps having programming experience fundamentally increases feelings of closeness to Alexa, seeing as students' \textit{prior} programming experience was moderately correlated with closeness. Furthermore, prior programming experience and human-likeness, as well as closeness and human-likeness were moderately correlated. One explanation could be that as students learned to program, they felt Alexa had human-like, logical reasoning, and thus felt closer to it (because of its human-like traits).

Students' perceptions of Alexa's friendliness and trustworthiness were strongly correlated, as well as trustworthiness and safeness, and to a lesser extent, intelligence and trustworthiness, friendliness and safeness, and closeness and trustworthiness. Although these correlations do not necessitate causation, it is important to consider the implications of potential causation when designing CAs. For instance, if a CA was purposefully designed to seem friendly and intelligent, users may associate this with trustworthiness and safeness, despite the potential for the CA to provide incorrect information (intentionally or not). Nevertheless, this could also provide positive opportunities, including how students may learn better if they feel a pedagogical agent is friendly and intelligent, and thus also trustworthy and safe. This is discussed in more depth below.

\subsection{Conceptions of AI and conversational AI}

From the pre-/post-test comparison of word frequency in responses describing AI (Fig. \ref{fig:wordcloud-ai}) and conversational AI (Fig. \ref{fig:wordcloud-conv-ai}), as well as the change in tag frequency analysis %
(Fig. \ref{fig:tags-conception}), students' conceptions seemed to shift towards more accurate understandings. For instance, the diction for describing AI seemed to shift towards computer-science-related terminology, including \textit{program}, \textit{learn}, and \textit{information}. This trend is consistent with other literature, in which students describe AI with more computer science vocabulary after developing AI projects \citep{learningml-rodriguez}. Furthermore, the emergence of the word \textit{learn} in post-test responses suggests a better understanding of AI systems' ability to adapt and update with training. For instance, one student's response described AI as ``a program that learns and uses the learning for other problems''. Furthermore, as shown in Fig. \ref{fig:tags-conception}, there was a relative increase in references to concepts from the Big AI Ideas \citep{big-ai-ideas-touretzky}, including \textit{Learning} and \textit{Representation and reasoning}, and a relative decrease in simplistic explanations of the AI acronym (e.g., ``AI is artificial intelligence'') and of how AI is ``like a human'', indicating better understanding.

In the conversational AI responses (Fig. \ref{fig:wordcloud-conv-ai}), \textit{human} remains the most frequent word in the pre- and post-test, suggesting student understanding of how human interaction is central to conversational AI's purpose. For the conversational AI descriptions shown in Fig. \ref{fig:tags-conception}, \textit{Learning} and the concept of natural language (NL) responses and understanding increased, indicating better understanding.
Furthermore, there was a relative decrease in simplistic explanations of conversational AI being ``like a robot'', mimicking humans, having pre-programmed responses, and being something that ``help[s] humans''. Despite these indications of better understanding, there was a slight relative increase in vague or shallow answers for both the descriptions of AI and conversational AI, and a relative decrease in the Big AI idea of representation and reasoning for conversational AI. Overall, however, it seemed as if students' conceptions improved through the workshops, especially considering the evidence for increased understanding of AI literacy concepts presented in \citep{appinv-convai-eaai}.

\subsection{Design Considerations}
Based on the results, we present design considerations for engaging students in %
learning experiences with CAs.

\subsubsection{Personification}
As shown in Tab. \ref{tab:ques-asked}, students asked Alexa many personal questions (e.g., ``Alexa, do you like Siri?'' and ``What's your favorite color?''), which would typically be asked of humans rather than computer systems. Alexa's often humorous responses (e.g., ``I like ultraviolet. It glows with everything'') could have contributed to students' perception of personified traits, like friendliness, intelligence and trustworthiness, which were all rated highly. As discussed, personified traits in CAs could play a role in effective teaching interventions \citep{avatar-digital-learning-schobel}, especially since feelings of closeness and trust can enhance human teaching and learning experiences \citep{teacher-child-closeness-adjustment-birch,gender-child-teacher-closeness-wolter,socioemotional-closeness-attachment-al-yagon}.

We recommend pedagogical CA developers \textbf{cautiously consider personification in their designs}. Although personification could engage students in effective learning experiences, it could also increase their feelings of trust disproportionately with the actual trustworthiness of the device. For example, students could perceive the device as \textit{always} providing unbiased, correct answers, despite AI systems often being biased \citep{ai-bias-roselli}. Thus, we further recommend considering transparency in CA design.

\subsubsection{Transparency}
Students also seemed to test the limits of Alexa, asking impossible or difficult questions as encapsulated by the \textit{Other} category in Tab. \ref{tab:ques-asked}. For example, students asked Alexa to turn itself off, to tell them all the (infinite) digits of $\pi$, and to provide the answer to $\frac{0}{0}$. These behaviors could be linked to trying to understand the system's inner workings. Thus, we recommend \textbf{developing CAs with the ability to explain themselves, and furthermore, provide transparency in terms of their abilities} (e.g., being able to explain AI bias). This is especially important when considering the correlations between CAs' friendliness and perceived trustworthiness, and students' potential increase in awareness of ignorance in how CAs work, as discussed above. This recommendation also aligns with other child-CA interaction research, which suggests designing transparent AI systems with respect to children's level of understanding \citep{popbots-robot-perceptions-randi}.

\subsubsection{Playfulness}
Similar to the behavior of ``testing'' Alexa described above, students asked Alexa playful questions like, ``How much wood would a wood chuck chuck if a wood chuck would chuck wood?'' and ``Are dragons real?''. These questions illustrate students'---even middle and high school students'---innate desire to play. Play can be hugely beneficial in learning environments, especially from a constructionist perspective \citep{constructionism-papert,playful-learning-rice}; thus, we recommend \textbf{considering playful learning experiences} when developing CAs. For example, in our study students had the opportunity to develop their own CA projects. Students came up with many different playful (as well as serious) ideas \citep{appinv-convai-eaai}. One very playful idea included a CA ``Meme Maker'', which according to the developer, ``help[ed] everyone get a quick laugh because as the old saying goes laughter is the best medicine''. This same student cited their favorite part of the workshop as ``improving [their] coding ability and learning more about [CAs]''.

\subsubsection{Utility}
Many student projects' purposes were to provide utility, with 34\% being mental and physical health-, 29\% being educational-, 21\% being productivity- and 8\% being accessibility-related CAs \citep{appinv-convai-eaai}. Utility was also reflected in students' interactions with Alexa, as \textit{Information updates} and \textit{Action commands} were the most common interactions reported. With students evidently being interested in CAs' utility, we recommend \textbf{designing CAs with useful features} to provide entry points to CA engagement and potential learning moments. For example, students might naturally engage with a CA in figuring out what the weather is like tomorrow, which would provide an opportunity to teach students about APIs and databases, and \textit{how} CAs provide such answers.

\section{Limitations and Future Work}
One limitation of this study includes its %
generalizability. We engaged middle and high school students in remote workshops in which they used MIT App Inventor to program Amazon Alexa; however, the results may not generalize to other environments or grade bands. Furthermore, since we held workshops on two different weeks with slight differences, %
this could have affected the results. Thus, future work may include larger follow-up studies with %
students in different grade bands in different environments.

There are also limitations associated with thematic analyses. For instance, we may have missed certain themes within the data, despite following the approach to analysis described in \citep{thematic-analysis-open-coding}. Furthermore, the amount of ideas presented by students in the pre- versus post-tests could have been influenced by the time of day each test was presented. Nevertheless, we believe the thematic data (as well as the word frequency data) are useful for exploratory, graphical analysis. Further research should statistically analyse students' conceptions of CAs and investigate how these conceptions affect the effectiveness of learning interventions.  %

\section{Conclusions}
Through the programming and learning intervention, students' perceptions of Alexa changed in how they viewed its intelligence and how close they felt to it, and students' conceptions tended towards describing AI systems using more computer science terminology and AI literacy concepts. Based on these results, we presented four design recommendations, including considering personification, transparency, playfulness and utility when designing CAs for engaging students in learning experiences. This study contributes to AI literacy research aiming to develop students' understanding of AI to be more accurate and healthy, ToAM research aiming to understand students' perception of AI, and CA research aiming to develop more useful, effective interactions.

\section{Selection and Participation of Children}

Children aged 11-18 (Mean=14.78, SD=1.91) were selected by their teachers to participate in the middle/high school workshops. Teachers were selected from those that responded to an Amazon Future Engineers call to Title I schools and signed a consent form. Selected students of the age of 18 were given similar student consent forms to sign, and those under the age of 18 were given assent forms and consent forms to be signed by their legal guardians before participating. The university's IRB approved the study protocol and consent/assent forms, which communicated how the data would be aggregated and anonymized. Given the wide age range, teachers assigned some of their older students to be mentors to younger students in case they fell behind.

\begin{acks}
We thank the teachers and students, volunteer facilitators, MIT App Inventor team, Personal Robots Group, and Amazon Future Engineer (AFE) members who made the workshops possible. Special thanks to Hal Abelson and Hilah Barbot. This work was funded by the AFE program and Hong Kong Jockey Club Charities Trust.
\end{acks}

\bibliographystyle{ACM-Reference-Format}
\bibliography{biblio}

\appendix

\end{document}